# Experimental Demonstration of Phase Modulation and Motion Sensing Using Graphene-Integrated Metasurfaces

Nima Dabidian[1], Shourya Dutta-Gupta[1], Iskandar Kholmanov[2], Feng Lu[3], Jongwon Lee[3], Kueifu Lai[1], Mingzhou Jin[3], Babak Fallahazad[3], Emanuel Tutuc[3], Mikhail A Belkin[3], Gennady Shvets[1]*

[1]Department of Physics and Center for Nano and Molecular Science and Technology, the University of Texas at Austin, Austin, Texas 78712, United States
[2]Department of Mechanical Engineering and Materials Science Program, The University of Texas at Austin, Austin, Texas 78712, United States
[3]Department of Electrical and Computer Engineering, Microelectronics Research Center, The University of Texas at Austin, 10100 Burnet Road, Austin, Texas 78758, USA.
*gena@physics.utexas.edu

**Abstract**: Plasmonic metasurfaces are able to modify the wavefront by altering the light intensity, phase and polarization state. Active plasmonic metasurfaces would allow dynamic modulation of the wavefront which give rise to interesting application such as beam-steering, holograms and tunable waveplates. Graphene is an interesting material with dynamic property which can be controlled by electrical gating at an ultra-fast speed. We use a graphene-integrated metasurface to induce a tunable phase change to the wavefront. The metasurface supports a Fano resonance which produces high-quality resonances around 7.7 microns. The phase change is measured using a Michleson interferometry setup. It is shown that the reflection phase can change up to 55 degrees. In particular the phase can change by 28° while the amplitude is nearly constant. The anisotropic optical response of the metasurface is used to modulate the ellipticity of the reflected wave in response to an incident field at $45°$. We show a proof of concept application of our system in potentially ultra-fast laser interferometry with sub-micron accuracy.

**Introduction:**

Metasurfaces are optical components that enhance the light-matter interaction and can control the flow of light. The flexibility in the design of these surfaces have brought unprecedented applications in biochemical sensing[1,2], wavefront engineering[3,4] and imaging[5]. These ultrathin devices are able to change the amplitude, polarization and phase of the light which are the necessary requirements to engineer the wavefront. Conventional optics uses propagation effect to gradually modify the light beam e.g. metamaterials with spatially varing indices can steer and control beam in applications such as optical cloaking[6,7] and superlens[8]. Metasurfaces on the other hand are able to introduce an abrupt change in optical properties which originates from the interaction of light with subwavelength antenna arrays. These arrays can have spatially varying optical response which provides great flexibility in molding the wavefront with applications in holograms[9–13], beam steering[14] and optical devices such as lenses[15], axicons[16] and waveplates[17].

The recent progress on active metasurfaces is promising to bring dynamics to these applications making them more even appealing for technological use. Graphene has emerged as an interesting active material due to the ultra-fast dynamics of electrical and optical properties [18,19] and broadband response [20]. This monolayer of carbon has been utilized in telecommunication[21–23] and electro-optical devices[24–28] across the spectrum. However, graphene can play an special role in the Mid-IR range due to limited choice of conventional optical elements and variety of technological applications. It has been shown that graphene tuning of Mid-IR light can be broad[29,30] and strong[31] due to its lossless plasmonic properties in this regime[32]. The response was shown to be in the order of nano-second limited only by the circuit delay[30]. Although there has been a few experimental demonstration of phase modulation in the terahertz regime using terahertz time-domain spectroscopy[24,33], to our knowledge there has been no experimental study on phase modulation using active metasrufaces in the Mid-IR. In this work we use a Michelson interferometric system for measuring the phase modulation induced by an active graphene metasurface. We show that it is possible to use the device for potentially ultra-fast electrical calibration of a laser interferometric setup. Such a system can be used for ultra-fast motion detection and surface topographic study. We finally demonstrate the tunable ellipticity of the reflected wave for a 45° incident polarization with potential applications in polarization-division multiplexing and ultrathin tunable waveplates.

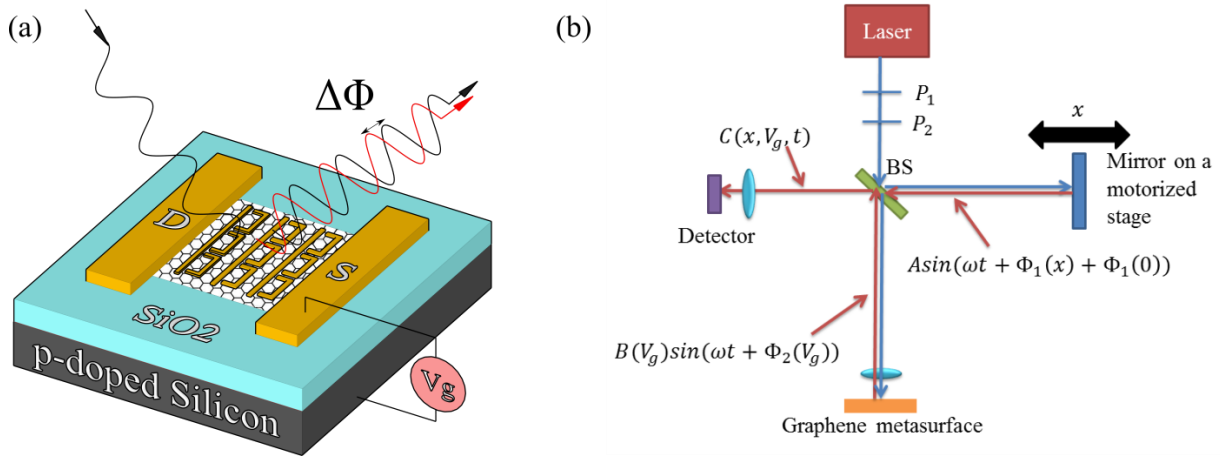

**Figure 1**: **(a)** Schematic for graphene induced modulation of the phase of scattered waves from a plasmonic metasurface. The metasurface is integrated with a single layer graphene that lies between drain and source contact. The phase modulation $\Delta\Phi$ is a function of the gate voltage $V_g$: the voltage between the backgate P-doped silicon and the source contact. The thickness of the $SiO_2$ spacer is 1 micron. **(b)** The schematic for Michelson interferometry setup with a quantum cascade laser source. The beam splitter divides the beam to two arm. One arm ends at a mirror which is positioned on a stage which moves by a clopped loop actuator with optical encoder. The phase of the wave reflected from the mirror (shown by the red arrow) $\Phi_1(x)$ is a function of mirror position x. On the other arm the laser beam is focused on a sample with a Fano metasurface integrated with graphene, the phase of the reflected beam $\Phi_2(V_g)$ is voltage-dependent. The IR detector reads the interference signal $C(x, V_g, t)$.

In an earlier communication, we showed 10 dB amplitude modulation, effectively switching the Mid-IR light by electrically controlling the charge concentration of single layer graphene (SLG) that is integrated to a plasmonic Fano metasurface[31]. In this work, we use a similar active metasurface, using SLG as active electro-optical material but focus on the phase modulation of the reflected waves as suggested in Fig. 1a where a backgate scheme is used to change graphene doping by an electrostatic gate voltage $V_g$. This in turn induces a gate-voltage dependent phase change $\Delta\Phi(V_g)$ in the wavefront of the waves reflected from the plasmonic metasurface. Using a Michelson interferometric setup as depicted in Fig. 1b, we measure the voltage dependent modulation of reflection phase. We use the plasmonic metasurface to enhance the interaction of electromagnetic fields with the single layer graphene (SLG). The unit cell of the metasurface is shown in Fig. 2a which consist of a continuous wire and a dipole. The metasurface exhibits electromagnetically induced transparency (EIT). This EIT is directly related to the Fano resonances which is defined by the interaction of a bright mode with broadband response with a narrowband dark mode[34]. The continuous wires of this structure mimic a dilute plasma and provide a broadband background. The structure supports a dark mode with quadrupolar charge distribution which manifests as a minimum in reflectivity. At the resonance, the current density of the dipole and the wire are anti-symmetric as shown in Fig. 2a. The interaction of light and graphene is proportional to $|E_t^2|$ [31,32] where $E_t$ is tangential electric field at the graphene surface. The field enhancement defined by $|E_t^2/E_{inc}^2|$ at the resonance is shown in Fig. 2a where $E_{inc}$ is the incident field. The metasurface is fabricated on top of SLG and an electrostatic gate voltage $V_g$ is applied between the silicon backgate and the graphene to change the carrier concentration $n$ of SLG which controls the optical conductivity of graphene. The reflectivity of the device was simulated using COMSOL multiphysics where graphene was modeled with a surface current: $J = \sigma_{SLG} E_t$ and random phase approximation[31,35] was used to model the optical conductivity of graphene $\sigma_{SLG}$ at different Fermi energies. The SLG charge concentration n is related to Fermi energy though $E_F = \hbar v_F \sqrt{\pi n}$ where $v_F = 1 * 10^8 \frac{cm}{s}$ is the Fermi velocity. The simulated reflectivity for Y-polarized light at different Fermi energies are shown in Fig. 2b. The graphene effect on the plasmonic metasurface are two-fold. (i) Spectral blue-shifting and (ii) spectral broadening of the resonance. Both of these effects are proportional to the tangential field enhancement $E_t^2/E_{inc}^2$ [31,32]. The backgate voltage $V_g$ determines the carrier density n according to $n = C_g \Delta V/e$, where $\Delta V = V_g - V_{CNP}$ is the potential deviation from the charge neutrality point (CNP) voltage $V_{CNP}$ that can be experimentally determined from the electrical measurements [20] as shown in Fig. 2d, and $C_g = \varepsilon/d$ is the gate capacitance per unit area; d and $\varepsilon$ are the thickness and electrostatic permittivity of the SiO$_2$ spacer. In order to characterize the electrical properties of graphene, we conduct an electrical transport measurement. The result is shown in Fig. 2d in terms of resistivity as a function of the gate voltage. The maximum resistance corresponds to charge neutrality point of graphene. For our sample $V_{CNP} \cong 40$ V. The graphene electrical properties such as contact resistance between graphene and the contacts ($R_c = 170\ \Omega$), electrical mobility in the hole-regime $\mu_h \cong 3600$ cm$^2$/Vs and the residual charges at the CNP point

$n_0 = 4.7e^{11}$ cm$^{-2}$ can be calculated from $R_{DS}(V_g)$ by a fitting procedure[36] as detailed in the method section. The red curve in Fig 2d demonstrates the fitted line in the hole regime. The carrier collisional time can be calculated from mobility from $\tau = \frac{\sigma_h \hbar^2}{2e^2 E_F} \approx 13 fs$ where $\sigma_h = n_h e \mu_h$ is the hole conductivity. The reflectivity of our sample is measured using laser spectroscopy as detailed in the method section. The results for different gate voltages are shown in Fig. 2e. The experiment is performed at normal incidence and for the light polarized along the dipole (Y-direction). The gate voltages in Fig 2e correspond to the Fermi energies in Fig. 2c and there is a good agreement between the measured and simulated reflectivity. An SEM picture of the metasurface is shown in the inset of Fig. 2f.

Our objective in this paper is to measure the graphene induced phase modulation of the electromagnetic waves and demonstrate proof of concept application of our device in motion sensing and waveplates.

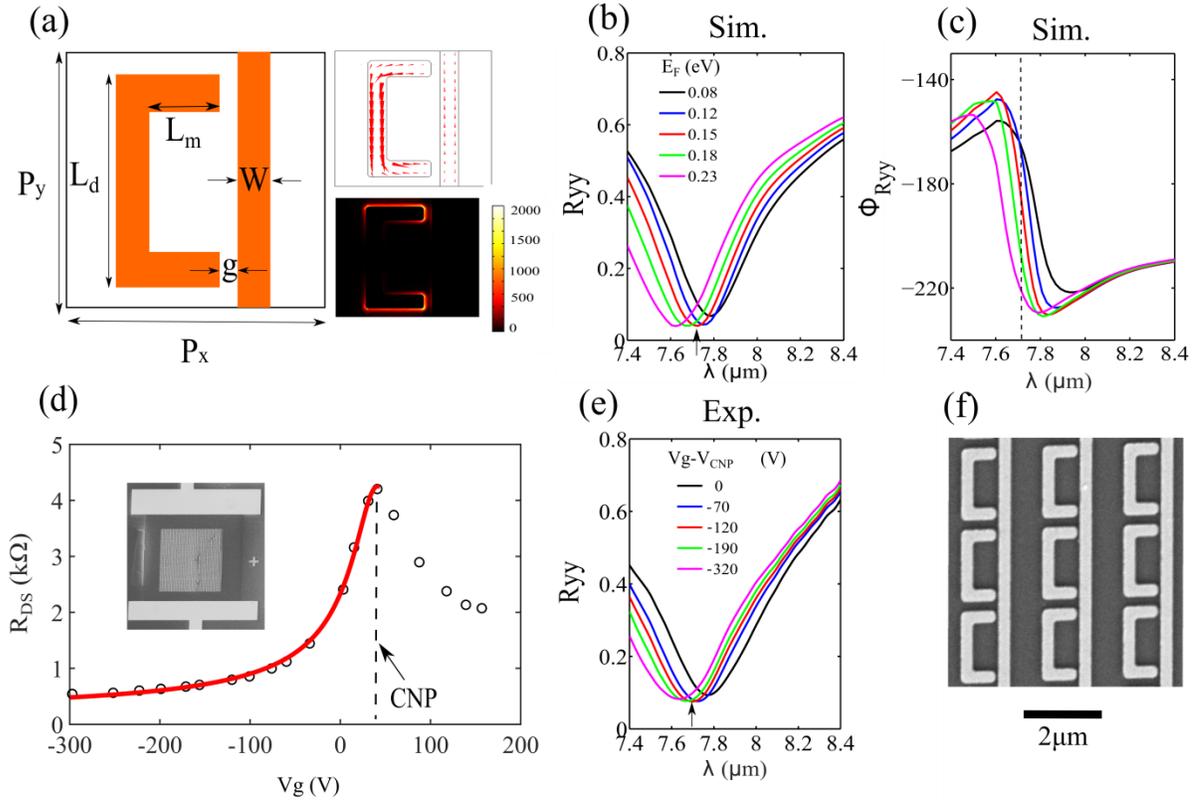

**Figure 2**: **(a)** The unit cell of the metasurface with the following dimensions: $g = 120\ nm, L_d = 1.8\ \mu m, L_m = 600\ nm$, the width of all the wires $W = 250\ nm$ and the periodicity in both directions $P_x = P_y = 2.1\ \mu m$. The current profile at the reflectivity minimum for the structure without graphene is shown on the top right. The current is plotted 5 nm above the SiO$_2$ surface. Near-field enhancement $|E_t^2/E_{inc}^2|$ at the Fano resonance frequency is plotted at the graphene plane (bottom right). **(b)** The simulated metasurface reflectivity spectrum for normal incidence with Y-polarized light. The colors represent different Fermi energies of the graphene. **(c)** The simulation results for reflection phase. Excitation parameters is similar to (b) **(d)** Electrical transport measurement for graphene: the resistance between the drain and the source contact ($R_{DS}$) as a function of gate voltage $V_g$ are shown by the black circles. The charge neutrality point (CNP) is shown by the dashed line at the maximum value of resistance at $V_g = 40\ V$. The inset is an SEM picture of the device showing the drain and source contacts on a graphene sheet. The metasurface is in the middle of graphene. The red curve is fitting $R_{DS}$ as a function of $V_g$ in the hole regime to find out the electrical properties of graphene as detailed in the method section. e) The measured reflectivity for y-polarized incident light at normal incidence. Different voltages have been color-coded according to their correspondent Fermi energy in (b). (f) An SEM picture of the metasurface. The black scale bar represents 2 micron.

**Interferometric measurement:**

We used a Michelson interferometry setup as shown in Fig. 1b to measure the phase modulation. The source is a quantum cascade laser. The laser light is split into two arms by a beam splitter (CaF$_2$ 2-8 um). On one arm, the beam is reflected from a mirror mounted on a motorized stage where the mirror motion is controlled by a closed-loop actuator with optical encoding capability. On the other arm, the graphene-integrated metasurface is mounted. The reflected wave (marked by red arrows) will combine on the beam-splitter and form interference which is detected by a MCT detector. The interfered wave can be written as equation 1:

$$C(x, V_g, t) = A\sin(\omega t + \Phi_1(x) + \Phi_1(0)) + B(V_g)\sin\left(\omega t + \Phi_2(V_g)\right) \tag{1}$$

Where A and $\Phi_1(x)$ are the amplitude and the position-dependent phase of the waves on the mirror arm. $B(V_g)$ and $\Phi_2(V_g)$ are the amplitude and the voltage-dependent phase of the reflected wave from the graphene-integrated metasurface. The difference between the path length of two arms brings in an initial phase represented by $\Phi_1(0)$ in equation 1. From equation 1, the time-averaged intensity is:

$$|C(x, V_g)|^2 = \frac{|A|^2}{2} + \frac{|B(V_g)|^2}{2} + AB(V_g)\cos(\Phi_1(x) + \Phi_1(0) - \Phi_2(V_g)) \tag{2}$$

$|C(x, V_g)|^2$ is the measured intensity for position $x$ of the mirror and graphene gate voltage $V_g$. Equation 2 consists of a DC offset and an interferometric part. To characterize $\Phi_2(V_g)$, we set the voltage to $V_g = 0$, move the mirror with constant step sizes, measuring the intensity after each step to observe a few full-wave interferometric oscillations in the measured amplitudes. We then return to the original position, change the voltage to $V_g = V_1$ and go forward for the same number of steps. By using a least-square fitting of the measured interferometric data to a cosine function with an offset: $a(V_g)\cos(bx + c(V_g)) + d(V_g)$ we can find the fitting parameters of $a(V_g), b, c(V_g)$ and $d(V_g)$ where $\Phi_1(x) = bx$ and $c(V_g) = \Phi_1(0) - \Phi_2(V_g)$. By comparing the fitted parameters for the two forward directions at to $V_g = 0$ and to $V_g = V_1$, the phase change can be calculated:

$$\Delta\Phi_2(V_g = 0, V_g = V_1) = [c_2(V_1) - c_2(0)] \tag{3}$$

and the phase modulation with respect to the CNP is defined as:

$$\Delta\Phi(V_g) = [\Phi_2(V_g) - \Phi_2(V_{CNP})] \tag{4}$$

The normalized interferometric signal can be defined from the fitted parameters:

$$I_N(x, V_g) = \frac{|C(x,V_g)|^2 - d(V_g)}{a(V_g)}. \tag{5}$$

In Fig. 3a $I_N(x, V_g = 0)$ and $I_N(x, V_g = -150)$ are shown with green and blue circles respectively. The fitted curves are demonstrated by the red and black solid lines in Fig. 3(a). The x-axis is the number of minimum step sizes. In principle the fitting parameter b is equal to $b = 4\pi/\lambda$ where $\lambda$ is the wavelength of light. However to characterize the actuator step size, we keep it as a fitting parameter. After fitting the interferometric signal in the two forward direction, we notice that $b_{f1}$ and $b_{f2}$ (parameter $b$ for the first and second forward direction) are roughly equal with 0.5 % maximum error for all of our experiments. To minimize the error we define $b = (b_{f1} + b_{f2})/2$ and use it in the fitting procedure mentioned earlier. In addition, from the fitting we find the minimum step size to be $\approx 68\ nm$.

The results of our interferometric experiments are shown in Fig. 3. In Fig. 3b, the fitted curves to $I_N(x, V_g)$ are demonstrated for three different Fermi energies at $\lambda_0 = 7.69\ \mu m$. The x-axis is the mirror movement normalized to the wavelength. The total phase shift for the Y-polarized light is about 55 degrees. To make sure our measurement results are reproducible we perform the phase measurement experiment 3 times. The measured phase change is shown in Fig. 3c which demonstrates the repeatability of our measurement system. In Fig. 3d the phase shift results of trial1 is shown in reflectivity-phase plane. Different Fermi energies are shown by different colors. The reflectivity in the colored region changes only about 10 % whereas the phase is changing by about 28 degrees.

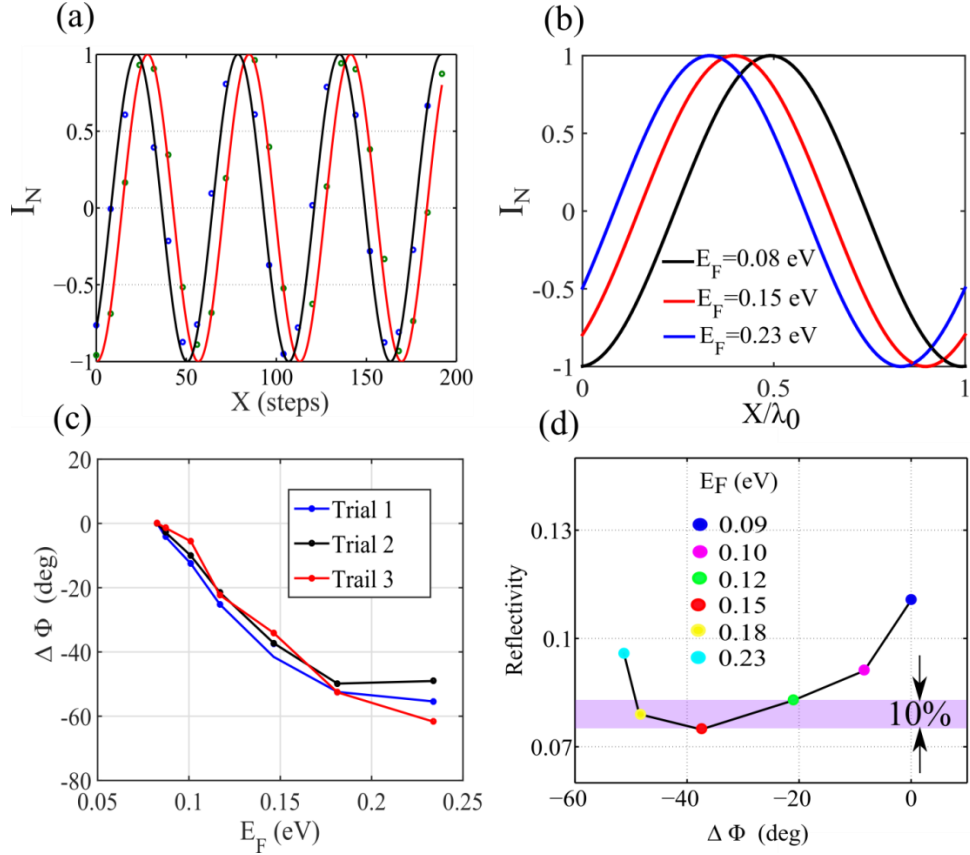

**Figure 3**: **(a)** The Interference for two different voltages: the x-axis shows mirror position in units number of minimum step sizes. The dots represent normalized interference data $I_N$ as defined by equation 5 for $V_g = 0$ (green) $V_g = -150$ (blue). These data are fitted to $cos(bx + c(V_g))$ where $b$ and $c(V_g)$ are the fitting parameters. All the experiments are run at $k_0 = 1300\ cm^{-1}$ ($\lambda_0 = 7.69\ \mu m$) **(b)** The normalized interference for 3 different Fermi energies. The x-axis is the mirror position change with respect to the origin normalized to the wavelength **(c)** The results of three independent interferometric measurement trials: Phase change $\Delta\Phi$ as a function of Fermi energy. The phase change is measured with respect to $V_g = V_{CNP}$ ($E_F = 0.08$ eV). The incident field was polarized along Y-direction. These three curves show the reproducibility of the experimental results **(d)** The resuls of trial1 is shown in the reflectivity-phase plane. The reflectivity changes about 10 % in the colored region while the phase is changing by about 28 degrees.

**Motion detection:**

So far we used a moving mirror with well-defined position to define the graphene induced phase modulation $\Delta\Phi(V_g)$ for our sample. In principle it is now possible to solve the inverse problem: define the motion of the mirror using the graphene sample with the known response function of $\Delta\Phi(V_g)$. Needless to say the mirror can be replaced by an arbitrary object with random motion. After calibrating the graphene sample we know the voltage dependent values of $a(V_g), b, c(V_g)$ and $d(V_g)$ where $\Delta\Phi(V_g)$ can be subsequently calculated from the fitting procedure. Now we are ready for solving the inverse problem. From equation 2 we know that the interfered signal depends on the initial phase $\Phi_1(0)$. Common laser interferometry systems have a mirror which moves mechanically to calibrate the system: calculate $\Phi_1(0)$ and determine the direction of the motion. However electro-optical modulation of phase such as in our graphene samples can calibrate the measurement much faster than the mechanical mirror motion permits. In what follows we show a proof of concept ultra-fast measurement of distance:

We start at a given mirror position and choose it as the reference point $x_0 = 0$. We apply different voltages and measure the voltage dependent intensity $|C(x_0, V_g)|^2$. We next use equation 5 to normalize the data in order to dissociate the interferometric intensity from voltage dependent amplitude $B(V_g)$. The result is $I_N(x_0, V_g)$. Next we move the mirror to other positions $x = x_1$ and $x = x_2$ and each time we follow same interferometric measurement and normalization procedure to calculate $I_N(x_1, V)$ and $I_N(x_2, V_g)$. Fig 4a shows the schematic for this experiment where mirror takes different positions. The results are shown in Fig. 4b where the dots represent the normalized interference $I_N(x_{0:2}, V_g)$ for voltage-dependent values of phase shift. Different colors correspond to different mirror position $x_{0:2}$. Next we fit these points to: $cos(c(V_g) + bx)$ where $c(V_g)$ and $b$ are known and $x$ is the fitting parameter. For simplicity we assume $x_0=0$ which implies $\Phi_1(0)=0$. We chose $x_0 = 0, x_1 = 544\ nm$ and $x_2 = 1088\ nm$. To make sure our procedure works, we performed the same experiment 5 times starting from random positions and each time we measured the position of the mirror $x_1$ and $x_2$ with respect to the origin $x_0 = 0$. The results are shown in Fig.4c. Different symbols represent different set of experiment. The solid blue line is our estimation based on the minimum step size ($68\ nm$) derived from the fitting. The results of the 5 different experiments are consistent with the reference blue line within 10 % accuracy.

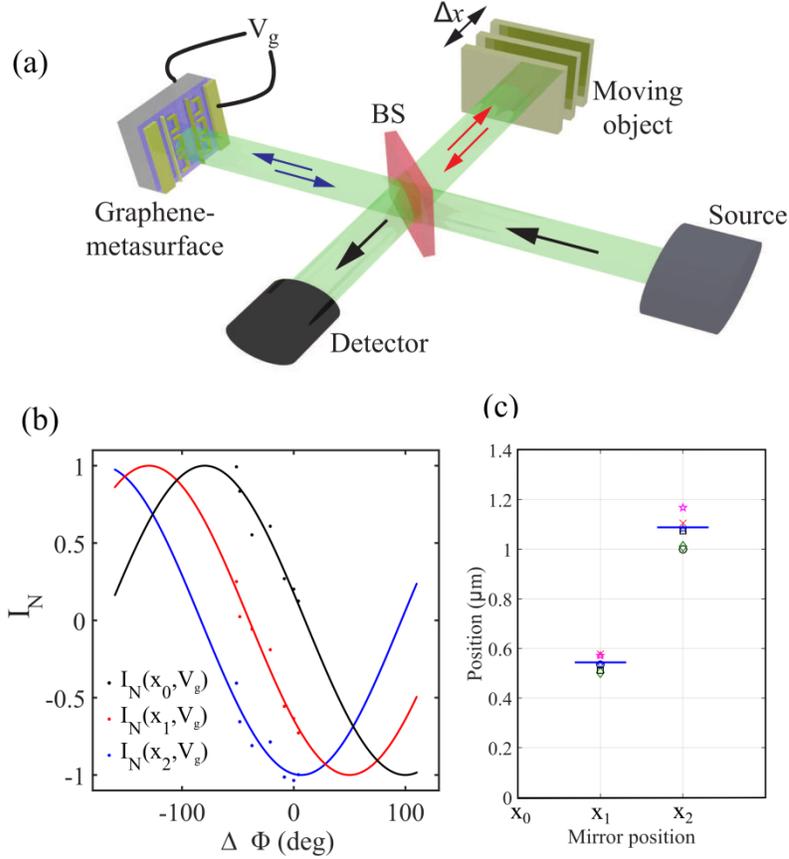

**Figure 4: (a)** The schematic for the motion detection setup **(b)** The normalized interferometric data (dots) for three different mirror position $x_0$, $x_1$ and $x_2$. The x-axis is $\Delta\Phi(V_g)$ which was determined through equation 4. The solid lines are the fitted curves to: $cos(c(V_g) + bx)$ where $x$ is the fitting parameters. The positions are directly calculated through the fitting. **(c)** The positions $x_1$ and $x_2$ calculated from the fitting are shown for five different trials assuming $x_0 = 0$ as the reference point.

**Application in ultrathin tunable waveplates:**

The metasurface is designed to have anisotropic response to X and Y polarization. While for Y-polarization there is a Fano resonance at around $\lambda = 7.7\mu m$, No mode exists for X-polarization at close proximity of this wavelength. In Fig 5a, the measured phase change for X and Y polarized light is shown at $\lambda = 7.69\mu m$. While for Y-polarization, the phase changes by about 50 degrees, for X-polarized light the change is only a few degrees. In Fig. 5b the measured reflectivity for 3 different Fermi energy is shown confirms small modulation for X-polarization. In Fig 5c, the simulation results for phase modulation are shown which are in good agreement with the experiment. The wavelength of our experiment is chosen to have the smallest reflectivity at $E_F = 0.15\ eV$ which is $\lambda = 7.69\mu m$ in experiment and $\lambda = 7.72\mu m$ in simulation as shown by the arrows in Fig 2b,e and dashed line in Fig. 2c. Now that the phase changes are characterized for both polarizations, one can calculate the elliptical polarization of the reflected wave in response to an incident beam at 45 degree as a function of Fermi energy. To depict the voltage dependent control over the polarization of the reflected light we plot the polarization ellipse for different gating voltages in Fig. 5d as detailed in the method section. It clearly shows that such a graphene-integrated metasurface can be used as an ultrathin tunable waveplates. The potentially ultra-fast modulation of ellipticity could be used in polarization-division data multiplexing to enhance the capacity of high-speed communication systems.

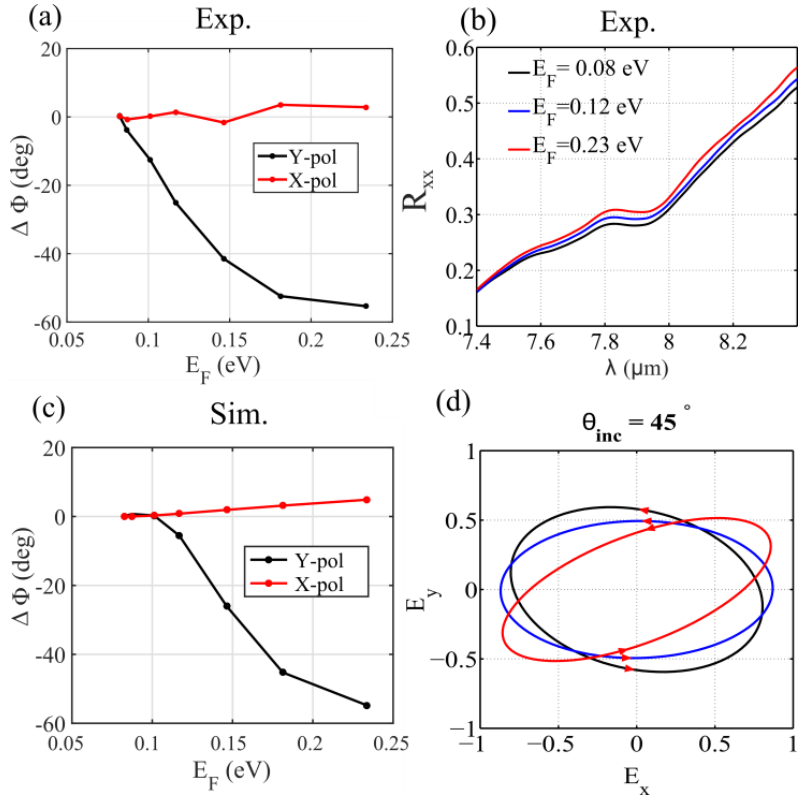

**Figure (5): (a)** The measured results for phase change as a function of Fermi energy for X and Y polarized incidence light. The phase is measured at the reflectivity minima for $E_F = 0.15$ eV which corresponds to $\lambda_0 = 7.69\ \mu m$ and $\lambda_0 = 7.72\ \mu m$ for simulation and experiment respectively **(b)** The measured reflectivity for X-polarized incident light for a few Fermi energies. **(c)** The simulation phase change for X and Y polarized incident light at different Fermi energies. The polarization ellipses of the reflected wave for 45° incidence as a function of Fermi energy. From simulations, at the Dirac point ($E_F = 0.08$ eV), $\Phi_{xy}(V_g = V_{CNP}) = 107°$.

**Conclusion:**

In summary, we showed that the measured phase modulation is close to the values predicted by simulations and the phase changes by about 28° while the amplitude is nearly constant which is relevant in sensing and inspires future works on beam steering. However the device has a large insertion loss ($\approx 10dB$) and requires large bias voltages. To achieve higher efficiency and higher phase change of $2\pi$ necessary for wavefront shaping, different designs should be considered.

## Methods:

**Sample Fabrication**: The process of sample fabrication was the following: as the first step, the SLG was grown on polycrystalline Cu foil using a CVD technique [37] and subsequently transferred from the Cu foil onto 1μm thick insulating (SiO$_2$) layer that was grown on a lightly doped silicon substrate [38] using wet thermal oxidation. Second, a 100μm × 100μm metasurface sample was fabricated on top of SLG with unit cell dimensions given in Fig. 2c using electron beam lithography (EBL). The thickness of the metasurface was 30nm (5nm of Cr and 25nm of Au). The inset of Fig. 2g, shows the SEM image of the Sample. Third, source and drain contacts (15nm Cr+85nm Au) were deposited on top of graphene on either sides of the metasurface samples using another EBL step. Lastly a gold contact (15nm Ni+85nm Au) was deposited on the backgate silicon. The fabrication was concluded by wirebonding to a chip carrier. An DC voltage was applied between the source contact and the silicon backgate to modulate graphene's carrier density as shown in Fig. 1.

**Graphene characterization and gating**: We used a parametric analyzer (Keithley 2450) for the current-voltage (I-V) measurement to characterize the SLG. In Fig. 2g the resistance between drain and source contacts $R_{DS}$ is shown as a function of gate voltage. This resistance can be written as [36]: $R_{DS} = R_c + R_g = R_c + N_{sq}/e \cdot \mu_h \cdot \sqrt{n_0^2 + n_{[V_g - V_{CNP}]}^2}$ where $N_{sq} = L_g/W_g$ with $L_g$ and $W_g$ being the length and the width of graphene channel. $\mu_h$ and $n_0$ represent the graphene hole mobility at room temperature and residual charge of graphene at the CNP point respectively. By fitting the experimentally measured resistance in Fig. 2g to $R_{DS}$, the fitting parameters $R_c = 170 \, \Omega$, $n_0 = 4.7e^{11} \, \text{cm}^{-2}$ and $\mu_h = 3600 \, \text{cm}^2/\text{Vs}$ can be derived. The charge neutrality point (CNP) $V_g = V_{CNP} = 40V$ corresponding to $n = n_0$ is identified by the maximum value of $R_{DS}(V_g)$. The slight p-doping of the SLG by the SiO2 substrate is inferred from $V_{CNP} > 0$. Due to the breakdown voltage of silicon dioxide at 0.5 GV/m, we vary the back gate voltage in the $-280V < V_g < 40V$ range using "Heathkit 500V PS-3" power supply. The holes' areal concentration can reach the maximum values of $n_h^{max} \approx 4.0e12 \, \text{cm}^{-2}$ for the peak gate voltage.

**Demonstration of polarization ellipses:** The ungated phase difference ($\Phi_{xy}(V_g = 0) = \Phi_x(V_g = 0) - \Phi_y(V_g = 0)$) for *x* and *y* polarized illuminations is fixed to 107° degrees, as estimated from numerical simulations. Additionally, we use the experimentally measured phase change (Fig. 3(c)) and reflectivity (Fig. 2(b)) as a function of voltage to calculate the total phase difference and relative intensities of *x* and *y* polarized components on gating. The polarization ellipse is then plotted using $E_x(V_g) = \sqrt{R_{xx}(V_g)}\exp(i\Phi_x(V_g) - i\omega t)$ and $E_y(V_g) = \sqrt{R_{yy}(V_g)}\exp(i\Phi_{xy}(V_g = 0) + i\Phi_y(V_g) - i\omega t)$, where $E_i(V_g)$, $R_i(V_g)$, and $\Phi_i(V_g)$ are the electric field, reflectivity and phase delay, respectively, on application of a gating voltage $V_g$ for $i$ polarization ($i=x, y$).

**Reflectivity measurement**: The setup shown in Fig 1b was used to measure the optical spectroscopy of the sample where the arm with the mirror is blocked by IR absorbing material and the only detected signal is from the graphene metasurface. The source is a quantum cascade laser (Daylight solution, MIRcat-1400). The laser was operated on pulsed mode with pulse repetitions rate of $250\ KHz$ with the pulse duration of $100\ ns$. A MCT detector was utilized for the measurement of intensity. The signal was amplified by a lock-in amplifier model (Stanford research systems SR844) with an integration time of $3\ ms$. A high numerical aperture NA=0.5 ZnSe lens was used as the objective.

**Interferometric measurement**: A Michelson interferometric setup as shown in Fig. 2a was used to measure the phase modulation. The mirror was mounted on a closed-loop actuator (Newport model 8310) equipped with an optical encoder which provides accurate step sizes. Each actuator jump consisted of 8 step sizes of roughly 68 nm. The integration time was set to be 100 ms. There is a 200ms delay time after an actuator jump till the intensity is measured. For each gate voltage the actuator moves for 25 jump in one direction.

**Numerical simulations**: A commercial finite elements solver COMSOL Multiphysics version 4.3b was used for simulations. The SLG was modeled using a surface current[32] $J_{SLG} = \sigma_{SLG} E_t$ where $E_t$ is the tangential electric field on the graphene plane.

## References:

(1) Wu, C.; Khanikaev, A. B.; Adato, R.; Arju, N.; Yanik, A. A.; Altug, H.; Shvets, G. Fano-Resonant Asymmetric Metamaterials for Ultrasensitive Spectroscopy and Identification of Molecular Monolayers. *Nat. Mater.* **2011**, *11* (1), 69–75.
(2) Adato, R.; Yanik, A. A.; Amsden, J. J.; Kaplan, D. L.; Omenetto, F. G.; Hong, M. K.; Erramilli, S.; Altug, H. Ultra-Sensitive Vibrational Spectroscopy of Protein Monolayers with Plasmonic Nanoantenna Arrays. *Proc. Natl. Acad. Sci.* **2009**, *106* (46), 19227–19232.
(3) Yu, N.; Capasso, F. Flat Optics with Designer Metasurfaces. *Nat. Mater.* **2014**, *13* (2), 139–150.
(4) Li, Z.; Yao, K.; Xia, F.; Shen, S.; Tian, J.; Liu, Y. Graphene Plasmonic Metasurfaces to Steer Infrared Light. *Sci. Rep.* **2015**, *5*, 12423.
(5) Liu, X.; Starr, T.; Starr, A. F.; Padilla, W. J. Infrared Spatial and Frequency Selective Metamaterial with Near-Unity Absorbance. *Phys. Rev. Lett.* **2010**, *104* (20), 207403.
(6) Shelby, R. A.; Smith, D. R.; Schultz, S. Experimental Verification of a Negative Index of Refraction. *Science* **2001**, *292* (5514), 77–79.
(7) Schurig, D.; Mock, J. J.; Justice, B. J.; Cummer, S. A.; Pendry, J. B.; Starr, A. F.; Smith, D. R. Metamaterial Electromagnetic Cloak at Microwave Frequencies. *Science* **2006**, *314* (5801), 977–980.
(8) Fang, N.; Lee, H.; Sun, C.; Zhang, X. Sub–Diffraction-Limited Optical Imaging with a Silver Superlens. *Science* **2005**, *308* (5721), 534–537.
(9) Larouche, S.; Tsai, Y.-J.; Tyler, T.; Jokerst, N. M.; Smith, D. R. Infrared Metamaterial Phase Holograms. *Nat. Mater.* **2012**, *11* (5), 450–454.

(10) Ni, X.; Kildishev, A. V.; Shalaev, V. M. Metasurface Holograms for Visible Light. *Nat. Commun.* **2013**, *4*.
(11) Tsai, Y.-J.; Larouche, S.; Tyler, T.; Llopis, A.; Royal, M.; Jokerst, N. M.; Smith, D. R. Arbitrary Birefringent Metamaterials for Holographic Optics at Λ = 1.55 Mm. *Opt. Express* **2013**, *21* (22), 26620–26630.
(12) Walther, B.; Helgert, C.; Rockstuhl, C.; Setzpfandt, F.; Eilenberger, F.; Kley, E.-B.; Lederer, F.; Tünnermann, A.; Pertsch, T. Spatial and Spectral Light Shaping with Metamaterials. *Adv. Mater.* **2012**, *24* (47), 6300–6304.
(13) Huang, L.; Chen, X.; Mühlenbernd, H.; Zhang, H.; Chen, S.; Bai, B.; Tan, Q.; Jin, G.; Cheah, K.-W.; Qiu, C.-W.; Li, J.; Zentgraf, T.; Zhang, S. Three-Dimensional Optical Holography Using a Plasmonic Metasurface. *Nat. Commun.* **2013**, *4*.
(14) Raghunathan, S. B.; Schouten, H. F.; Ubachs, W.; Kim, B. E.; Gan, C. H.; Visser, T. D. Dynamic Beam Steering from a Subwavelength Slit by Selective Excitation of Guided Modes. *Phys. Rev. Lett.* **2013**, *111* (15), 153901.
(15) Aieta, F.; Genevet, P.; Kats, M. A.; Yu, N.; Blanchard, R.; Gaburro, Z.; Capasso, F. Aberration-Free Ultrathin Flat Lenses and Axicons at Telecom Wavelengths Based on Plasmonic Metasurfaces. *Nano Lett.* **2012**, *12* (9), 4932–4936.
(16) MCLEOD, J. H. The Axicon: A New Type of Optical Element. *J. Opt. Soc. Am.* **1954**, *44* (8), 592–592.
(17) Yu, N.; Aieta, F.; Genevet, P.; Kats, M. A.; Gaburro, Z.; Capasso, F. A Broadband, Background-Free Quarter-Wave Plate Based on Plasmonic Metasurfaces. *Nano Lett.* **2012**, *12* (12), 6328–6333.
(18) Tielrooij, K. J.; Piatkowski, L.; Massicotte, M.; Woessner, A.; Ma, Q.; Lee, Y.; Myhro, K. S.; Lau, C. N.; Jarillo-Herrero, P.; Hulst, N. F. van; Koppens, F. H. L. Generation of Photovoltage in Graphene on a Femtosecond Timescale through Efficient Carrier Heating. *Nat. Nanotechnol.* **2015**, *10* (5), 437–443.
(19) Xia, F.; Mueller, T.; Lin, Y.; Valdes-Garcia, A.; Avouris, P. Ultrafast Graphene Photodetector. *Nat. Nanotechnol.* **2009**, *4* (12), 839–843.
(20) Novoselov, K. S.; Geim, A. K.; Morozov, S. V.; Jiang, D.; Katsnelson, M. I.; Grigorieva, I. V.; Dubonos, S. V.; Firsov, A. A. Two-Dimensional Gas of Massless Dirac Fermions in Graphene. *Nature* **2005**, *438* (7065), 197–200.
(21) Mohsin, M.; Neumaier, D.; Schall, D.; Otto, M.; Matheisen, C.; Lena Giesecke, A.; Sagade, A. A.; Kurz, H. Experimental Verification of Electro-Refractive Phase Modulation in Graphene. *Sci. Rep.* **2015**, *5*, 10967.
(22) Midrio, M.; Galli, P.; Romagnoli, M.; Kimerling, L. C.; Michel, J. Graphene-Based Optical Phase Modulation of Waveguide Transverse Electric Modes. *Photonics Res.* **2014**, *2* (3), A34.
(23) Zhou, F.; Hao, R.; Jin, X.-F.; Zhang, X.-M.; Li, E.-P. A Graphene-Enhanced Fiber-Optic Phase Modulator With Large Linear Dynamic Range. *IEEE Photonics Technol. Lett.* **2014**, *26* (18), 1867–1870.
(24) Lee, S. H.; Choi, M.; Kim, T.-T.; Lee, S.; Liu, M.; Yin, X.; Choi, H. K.; Lee, S. S.; Choi, C.-G.; Choi, S.-Y.; Zhang, X.; Min, B. Switching Terahertz Waves with Gate-Controlled Active Graphene Metamaterials. *Nat. Mater.* **2012**, *11* (11), 936–941.
(25) Emani, N. K.; Chung, T.-F.; Kildishev, A. V.; Shalaev, V. M.; Chen, Y. P.; Boltasseva, A. Electrical Modulation of Fano Resonance in Plasmonic Nanostructures Using Graphene. *Nano Lett.* **2013**.
(26) Sensale-Rodriguez, B.; Yan, R.; Kelly, M. M.; Fang, T.; Tahy, K.; Hwang, W. S.; Jena, D.; Liu, L.; Xing, H. G. Broadband Graphene Terahertz Modulators Enabled by Intraband Transitions. *Nat. Commun.* **2012**, *3*, 780.
(27) Jang, M. S.; Brar, V. W.; Sherrott, M. C.; Lopez, J. J.; Kim, L.; Kim, S.; Choi, M.; Atwater, H. A. Tunable Large Resonant Absorption in a Midinfrared Graphene Salisbury Screen. *Phys. Rev. B* **2014**, *90* (16), 165409.

(28) Brar, V. W.; Sherrott, M. C.; Jang, M. S.; Kim, S.; Kim, L.; Choi, M.; Sweatlock, L. A.; Atwater, H. A. Electronic Modulation of Infrared Radiation in Graphene Plasmonic Resonators. *Nat. Commun.* **2015**, *6*, 7032.
(29) Yao, Y.; Kats, M. A.; Genevet, P.; Yu, N.; Song, Y.; Kong, J.; Capasso, F. Broad Electrical Tuning of Graphene-Loaded Plasmonic Antennas. *Nano Lett.* **2013**, *13* (3), 1257–1264.
(30) Yao, Y.; Kats, M. A.; Shankar, R.; Song, Y.; Kong, J.; Loncar, M.; Capasso, F. Wide Wavelength Tuning of Optical Antennas on Graphene with Nanosecond Response Time. *Nano Lett.* **2014**, *14* (1), 214–219.
(31) Dabidian, N.; Kholmanov, I.; Khanikaev, A. B.; Tatar, K.; Trendafilov, S.; Mousavi, S. H.; Magnuson, C.; Ruoff, R. S.; Shvets, G. Electrical Switching of Infrared Light Using Graphene Integration with Plasmonic Fano Resonant Metasurfaces. *ACS Photonics* **2015**, *2* (2), 216–227.
(32) Mousavi, S. H.; Kholmanov, I.; Alici, K. B.; Purtseladze, D.; Arju, N.; Tatar, K.; Fozdar, D. Y.; Suk, J. W.; Hao, Y.; Khanikaev, A. B.; Ruoff, R. S.; Shvets, G. Inductive Tuning of Fano-Resonant Metasurfaces Using Plasmonic Response of Graphene in the Mid-Infrared. *Nano Lett.* **2013**, *13* (3), 1111–1117.
(33) Chen, H.-T.; Padilla, W. J.; Cich, M. J.; Azad, A. K.; Averitt, R. D.; Taylor, A. J. A Metamaterial Solid-State Terahertz Phase Modulator. *Nat. Photonics* **2009**, *3* (3), 148–151.
(34) Khanikaev, A. B.; Wu, C.; Shvets, G. Fano-Resonant Metamaterials and Their Applications. *Nanophotonics* **2013**, *2* (4), 247–264.
(35) Falkovsky, L. A.; Pershoguba, S. S. Optical Far-Infrared Properties of a Graphene Monolayer and Multilayer. *Phys. Rev. B* **2007**, *76* (15), 153410.
(36) Kim, S.; Nah, J.; Jo, I.; Shahrjerdi, D.; Colombo, L.; Yao, Z.; Tutuc, E.; Banerjee, S. K. Realization of a High Mobility Dual-Gated Graphene Field-Effect Transistor with Al2O3 Dielectric. *Appl. Phys. Lett.* **2009**, *94* (6), 062107.
(37) Li, X.; Zhu, Y.; Cai, W.; Borysiak, M.; Han, B.; Chen, D.; Piner, R. D.; Colombo, L.; Ruoff, R. S. Transfer of Large-Area Graphene Films for High-Performance Transparent Conductive Electrodes. *Nano Lett.* **2009**, *9* (12), 4359–4363.
(38) Kholmanov, I. N.; Magnuson, C. W.; Aliev, A. E.; Li, H.; Zhang, B.; Suk, J. W.; Zhang, L. L.; Peng, E.; Mousavi, S. H.; Khanikaev, A. B.; Piner, R.; Shvets, G.; Ruoff, R. S. Improved Electrical Conductivity of Graphene Films Integrated with Metal Nanowires. *Nano Lett.* **2012**, *12* (11), 5679–5683.